%% file: main.tex
\documentclass[letterpaper, 10 pt, conference]{ieeeconf}

\IEEEoverridecommandlockouts   %
\usepackage[utf8]{inputenc}
\usepackage{amsmath,amssymb,amsfonts}
\usepackage{algorithmic}
\usepackage{algorithm}
\usepackage{graphicx}
\graphicspath{{./}}
\usepackage{textcomp}
\usepackage{xcolor}
\usepackage{booktabs}
\usepackage{multirow}
\usepackage{balance}
\usepackage{placeins}
\usepackage{needspace}
\usepackage[hidelinks]{hyperref}
\usepackage{bm}
\usepackage{tikz}
\usetikzlibrary{arrows.meta,positioning,calc}

\usepackage{iftex}
\ifPDFTeX\else
  \usepackage{fontspec}
  \IfFontExistsTF{Times New Roman}{\setmainfont{Times New Roman}}{}
\fi

\newcommand{\R}{\mathbb{R}}
\newcommand{\Vdot}{\dot{V}}
\newcommand{\norm}[1]{\|#1\|}
\newcommand{\method}{SLyM}

\title{\LARGE \bf SLyM: LLM-Driven Symbolic Lyapunov Function Discovery via Structure Proposal and Coefficient Optimization}
\author{Zihao Fu\authorrefmark{1}, Xingchen Li\authorrefmark{1}, Keyou You%
\thanks{*These authors contributed equally to this work.}%
\thanks{This work was supported by National Natural Science Foundation of China (62325305). (Corresponding author: Keyou You)}%
\thanks{Zihao Fu, Xingchen Li, and Keyou You are with the Department of Automation, and Beijing National Research Center for Info. Sci. \& Tech. (BNRist), Tsinghua University, Beijing 100084, China. (e-mail: \{fuzh24,lixc21\}@mails.tsinghua.edu.cn, youky@tsinghua.edu.cn)}%
}

\begin{document}

\maketitle
\thispagestyle{empty}
\pagestyle{empty}

\input{body}

\end{document}

%% file: body.tex
\newtheorem{definitionX}{Definition}
\newtheorem{propositionX}{Proposition}
\newenvironment{definition}[1][]{%
  \par\smallskip
  \refstepcounter{definitionX}%
  \noindent\textit{Definition~\thedefinitionX
    \if\relax\detokenize{#1}\relax\else\ (#1)\fi.}\ %
  \normalfont\setlength{\parindent}{0pt}%
}{\par\smallskip}
\newenvironment{proposition}[1][]{%
  \if\relax\detokenize{#1}\relax
    \begin{propositionX}\normalfont
  \else
    \begin{propositionX}[#1]\normalfont
  \fi
}{\end{propositionX}}
\newenvironment{pf}{\par\noindent\emph{Proof.}\ }{\hfill$\square$\par}
\definecolor{axblue}{RGB}{48,111,199}
\definecolor{axred}{RGB}{218,56,50}
\definecolor{axyellow}{RGB}{246,194,74}
\definecolor{axgreen}{RGB}{67,147,143}
\definecolor{axterra}{RGB}{186,110,74}

\renewcommand{\topfraction}{0.80}
\renewcommand{\bottomfraction}{0.45}
\renewcommand{\textfraction}{0.18}
\renewcommand{\floatpagefraction}{0.70}
\setcounter{topnumber}{2}
\setcounter{bottomnumber}{1}
\setcounter{totalnumber}{3}

\setlength{\textfloatsep}{8pt plus 2pt minus 2pt}
\setlength{\floatsep}{6pt plus 2pt minus 2pt}
\setlength{\intextsep}{6pt plus 2pt minus 2pt}
\setlength{\dbltextfloatsep}{8pt plus 2pt minus 2pt}
\setlength{\dblfloatsep}{6pt plus 2pt minus 2pt}

\begin{abstract}
Lyapunov functions are the cornerstone of stability analysis for dynamical systems, yet synthesizing a symbolic form remains difficult as no universal constructive algorithm exists. To address this challenge, we introduce \method{}, a framework that harnesses the knowledge and reasoning of large language models (LLMs) to discover symbolic Lyapunov functions. To avoid charging the LLM with ill-suited numerical generation, \method{} decomposes the construction into \emph{structure proposal} and \emph{coefficient optimization}: the LLM draws on its prior knowledge of Lyapunov constructions to propose the function structure, whose coefficients then follow from a convex program. Local-search structures are folded back into supervision, extending coverage to systems missed by the initial proposer. Across all benchmarks, \method{} consistently outperforms the evaluated neural, sum-of-squares, and generative baselines.
\end{abstract}

\section{Introduction}\label{sec:intro}
Lyapunov functions provide the foundational framework for stability analysis of dynamical systems. For a given system, once a Lyapunov function is constructed, its stability can be validated and the result provides key theoretical support for control design~\cite{lyapunov1892,khalil2002nonlinear,dawson2023survey}. However, synthesizing such a function remains notoriously difficult: no universal constructive algorithm exists, and classical converse theorems guarantee existence without providing any systematic means to determine a symbolic form~\cite{khalil2002nonlinear}.

Classical practice relies on parameterized families or system-specific constructions---such as energies, linearizations, or backstepping~\cite{khalil2002nonlinear,strogatz2015nonlinear}---which are confined to particular structures and posited anew for each system. Computational tools automate this search over prescribed finite-dimensional spaces: sum-of-squares (SOS) programming~\cite{prajna2002sostools} and fixed-basis methods~\cite{sankaranarayanan2013handelman,bensassi2016linear} fix the basis in advance and optimize only coefficients. Yet fixing the representation beforehand severely limits expressiveness: even globally asymptotically stable polynomial systems need not admit any polynomial Lyapunov function~\cite{ahmadi2011globally}, while non-polynomial dynamics remain largely inaccessible.

Rather than fixing a basis a priori, neural Lyapunov methods parameterize Lyapunov functions as neural networks~\cite{chang2019neural,abate2020formal}. Despite their expressive flexibility, these models yield implicit black-box representations whose verification poses severe scalability bottlenecks---frequently causing satisfiability modulo theories (SMT) timeouts even on modest 4--6-dimensional systems~\cite{dawson2023survey}. In contrast, compact symbolic Lyapunov functions can ease verification and provide explicit expressions for stability analysis~\cite{khalil2002nonlinear}.

To discover such symbolic functions, large language models (LLMs) offer an expressive, data-driven prior. Pretrained on extensive mathematical and engineering literature, an LLM internalizes canonical Lyapunov constructions---from positive-definite quadratic backbones to energy-like potentials and cross-terms that neutralize nonlinear drift. Furthermore, by exploiting structural cues in the dynamics, the model can suggest candidate terms tailored to the system, spanning polynomial and non-polynomial forms alike. The success of LLMs in mathematical discovery further underscores the promise of this direction~\cite{yang2023leandojo,romera2024funsearch}.

We therefore introduce \method{} (Symbolic Lyapunov functions with LLM), an LLM-driven framework that decomposes Lyapunov-function construction into two subproblems: \emph{structure proposal}, which determines candidate symbolic terms, and \emph{coefficient optimization}, which computes their numerical weights. This separation matters: recent symbolic methods emit complete expressions with embedded constants~\cite{alfarano2024global,zou2025analytical}, burdening generative models with ill-suited numerical fitting. In contrast, \method{} decouples the two: conditioned on system dynamics, the LLM leverages its symbolic prior to propose a coefficient-free structure beyond rigid fixed bases, while continuous coefficients are obtained by convex optimization.

To better adapt the LLM to this task, we adopt a two-stage training strategy. First, supervised fine-tuning on successful examples teaches the LLM the mapping from system dynamics to valid structures. Second, structures discovered by local structure search are folded back into the training loop, enabling the model to internalize search-time discoveries and propose viable structures more directly on unseen systems.

Our contributions are threefold: First, we propose \method{}, an LLM-driven framework that decomposes symbolic Lyapunov discovery into structure proposal and coefficient optimization, choosing system-dependent structures from a library of symbolic building blocks while avoiding ill-suited numerical generation. Second, we develop a search-augmented training scheme that folds search-recovered structures back into supervision, extending coverage to systems missed by the initial proposer. Third, \method{} consistently outperforms the evaluated neural, sum-of-squares, and generative baselines across extensive benchmarks. Furthermore, on systems drawn from the control literature, it finds accepted candidates different from the reference functions in the cited publications.

\section{Preliminaries}\label{sec:preliminaries}

\subsection{Dynamical Systems and Lyapunov Functions}\label{sec:prelim}

{
Consider the autonomous system
\begin{equation}\label{eq:sys}
\dot{x}=f(x),\qquad f\in C^{1}(U,\R^{n}),\qquad U\subseteq\R^{n}\ \text{open},
\end{equation}
where $n$ is the state dimension. We assume, without loss of generality, that
the equilibrium of interest is at the origin, so that $f(0)=0$. Let
$\Omega\subset U$ be a compact set with $0\in\operatorname{int}\Omega$. Given
$f$ and $\Omega$, we seek a symbolic Lyapunov function on $\Omega$. The
standard Lyapunov condition for this system is stated as follows.
\begin{definition}[Lyapunov function~\cite{khalil2002nonlinear}]\label{def:lyap}A function $V\in C^{1}(U,\R)$ is a \emph{Lyapunov function} for the equilibrium
$x=0$ of \eqref{eq:sys} on $\Omega$ if $V$ is positive definite on $\Omega$
and its derivative along system trajectories is negative semidefinite, i.e.,
\begin{equation}\label{eq:pd}
V(0)=0,\qquad V(x)>0,\quad \forall x\in\Omega\setminus\{0\},
\end{equation}
\begin{equation}\label{eq:nd}
\Vdot(x):=\nabla V(x)^{\!\top} f(x) \le 0,\quad \forall x\in\Omega.
\end{equation}
\end{definition}
}
{
\noindent
By Lyapunov's stability theorem, \eqref{eq:pd}--\eqref{eq:nd} imply that the origin
is stable. To illustrate these conditions, consider
the following non-polynomial planar system:
}
\begin{equation}\label{eq:levels_sys}
\left\{\begin{aligned}
\dot x_1&=-7x_1-x_2^{2}e^{x_1+x_2},\\
\dot x_2&=-6x_1^{2}x_2^{3}+5x_2e^{x_2}\sin x_1.
\end{aligned}\right.
\end{equation}
As shown in Fig.~\ref{fig:levels}, condition \eqref{eq:pd} requires $V$ to form a bowl-shaped surface with a unique minimum at the origin (Fig.~\ref{fig:levels}a), while \eqref{eq:nd} ensures that system trajectories never cross outward across its nested level sets (Fig.~\ref{fig:levels}b).

\begin{figure}[t]
\centering
\includegraphics[width=\columnwidth]{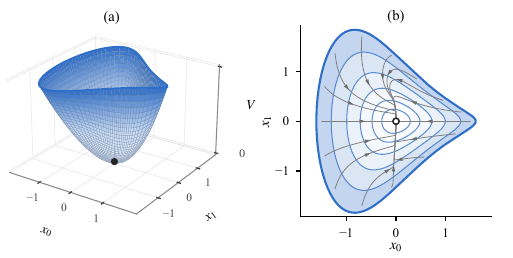}
\caption{A candidate recovered by \method{} for system \eqref{eq:levels_sys}: $V=90x_1^2+x_2^2(75e^{2x_1}+36e^{x_1}+20)$. \emph{(a)} Surface of $V$. \emph{(b)} Level sets and trajectories inside the plotted sublevel region $V\le225$.}
\label{fig:levels}
\end{figure}

\subsection{Structured Lyapunov Discovery}\label{sec:struct}

{
Definition~\ref{def:lyap} specifies the conditions a Lyapunov function must satisfy
without prescribing a particular functional form. A common approach to synthesis
is to fix a parametric family of candidate functions and solve for its unknown
coefficients~\cite{sankaranarayanan2013handelman,bensassi2016linear}. In these
methods, the structure is chosen in advance. We follow this
structure--coefficient decomposition while making the structure system
dependent: \emph{structure proposal} selects a compact set of component
functions whose linear combination forms a candidate Lyapunov function, and
\emph{coefficient optimization} determines the corresponding numerical
weights. An LLM performs the first task, while a convex program performs the
second.
}

{
To implement this decomposition, we define a library
$\mathcal B\subset C^{1}(U,\R)$ of candidate building blocks. Each
$b\in\mathcal B$ is formed from the state variables using multiplication,
positive integer powers up to $p_{\max}$, and composition with $\exp$, $\sin$,
and $\cos$. We write each term without a scalar prefactor and store that
prefactor in the coefficient vector; for example, $3x_1^2$ is represented by
the term $x_1^2$ with coefficient $3$. After canonicalization, a candidate is
specified by a subset $S$ of these terms and its coefficient vector $c$:
}
\begin{equation}\label{eq:struct}
\begin{aligned}
S&=\{b_1,\dots,b_m\}\subset\mathcal B, &&1\le m\le m_{\max},\\
c&=[c_1,\dots,c_m]^{\top}\in\R^{m},
\end{aligned}
\end{equation}
where $m=\lvert S\rvert$ and $c_j$ is the coefficient of $b_j$.

{
For example,
\[
V(x)=\tfrac16x_1(e^{x_1}-1)+x_2^{2}
=\tfrac16 x_1e^{x_1}-\tfrac16 x_1+x_2^{2}
\]
corresponds to the structure $S=\{x_1e^{x_1},x_1,x_2^{2}\}$ and the
coefficient vector $c=(\tfrac16,-\tfrac16,1)$.
}

For a given structure, the candidate and its Lie derivative are linear in $c$:
\begin{equation}\label{eq:cand}
\begin{aligned}
V_{S,c}(x)&=\sum_{j=1}^{m}c_jb_j(x)=c^{\top}\Phi_S(x),\\[-1pt]
\Vdot_{S,c}(x)&=\sum_{j=1}^{m}c_j\nabla b_j(x)^{\!\top}f(x)
                 =c^{\top}\dot\Phi_S(x),
\end{aligned}
\end{equation}
where $\Phi_S$ collects the selected terms and $\dot\Phi_S$ is its derivative
along the vector field.
{%
For fixed $S$, coefficient optimization is convex because both expressions are
linear in $c$; structure proposal is the remaining combinatorial search. We
formalize the resulting nested optimization by giving
\eqref{eq:pd}--\eqref{eq:nd} a quantitative form with margin $\gamma>0$, letting
$u,v\ge0$ bound their violations:
}
\begin{equation}\label{eq:nestedsearch}
\begin{gathered}
\inf_{\substack{S\subset\mathcal B\\ m\le m_{\max}}}\ J_{\gamma}(S):=\!\!\min_{\substack{\norm{c}_\infty\le1\\ u,v\ge0}}\ u+v\\[1pt]
\text{s.t.}\quad
\begin{aligned}
V_{S,c}(x)-V_{S,c}(0)&\ge \gamma\norm{x}^{2}-u,\\[-1pt]
\Vdot_{S,c}(x)&\le-\gamma\norm{x}^{2}+v,
\end{aligned}
\quad\forall x\in\Omega.
\end{gathered}
\end{equation}

{
For a fixed $S$, \eqref{eq:nestedsearch} minimizes the worst-case violations
$u+v$ over $\Omega$. The bound $\norm{c}_\infty\le1$ prevents trivial scaling,
and $\gamma>0$ prevents $V\equiv0$ from attaining zero loss. Thus, $J_{\gamma}(S)=0$ yields a
candidate satisfying the conditions in Definition~\ref{def:lyap} with
$\Vdot_{S,c}\le-\gamma\norm{x}^{2}$ after subtracting its value at the
origin. These margin conditions are sufficient, not necessary.
Section~\ref{sec:details} describes how
\method{} searches over $S$ and relaxes the coefficient problem over
$\Omega$ to a finite-sample convex program.
}

\section{\method{}: Structure Proposal and Coefficient Optimization}\label{sec:details}

\begin{figure}[t]
\centering
\input{fig_pipeline.tex}
\caption{\method{} follows the two-level view in \eqref{eq:nestedsearch}: structure proposal explores the symbolic structure with an LLM proposer and a local structure search, coefficient optimization computes the numerical weights, and candidate validation evaluates the resulting symbolic function. Stage~2 trains on structures recovered by local search after initial proposals fail.}
\label{fig:pipeline}
\end{figure}
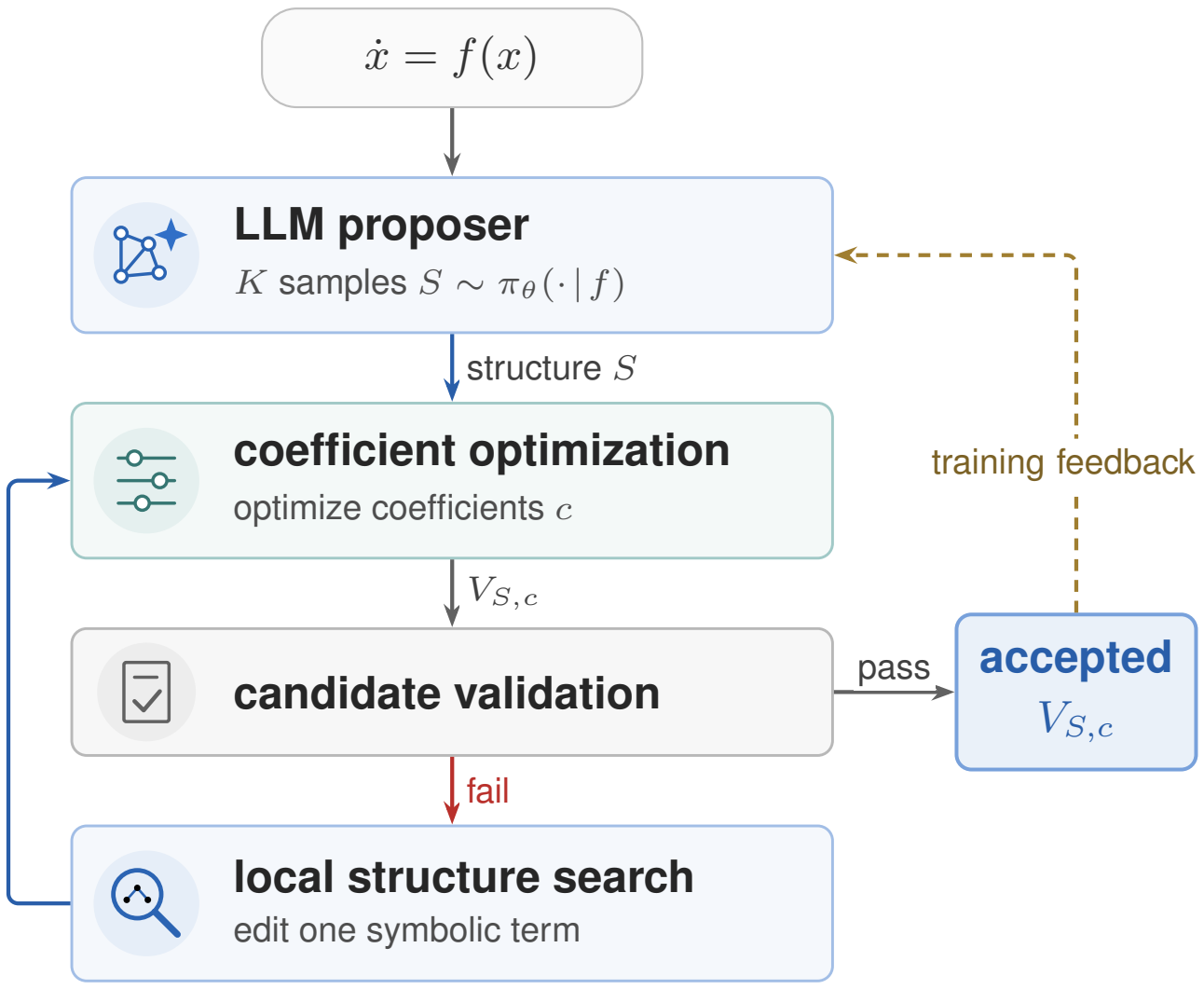

{
For a fixed structure, \eqref{eq:nestedsearch} is convex in $c$;
the difficulty lies in choosing $S$. Admissible structures grow
combinatorially with the library, the structures making
\eqref{eq:pd}--\eqref{eq:nd} feasible vary across systems, and each costs a
coefficient fit. \method{} therefore assigns the
combinatorial choice to a learned proposer and the continuous one to
optimization (Fig.~\ref{fig:pipeline}): an LLM proposes structures, refined by
local search when needed (Section~\ref{sec:structsearch}),
search-recovered structures are fed back into training
(Section~\ref{sec:training}), and coefficients follow from a finite-sample
convex program before validation (Section~\ref{sec:coeff}).
}

\subsection{Structure Proposal}\label{sec:structsearch}

{
The first step in the pipeline is to propose a compact structure adapted to the
dynamics, whose form often suggests suitable terms:
nonlinearities in $f$, such as exponential or trigonometric factors, tend to
call for matching terms in $V$. A fixed basis may miss interactions needed by a particular system,
whereas a broad library makes search and verification expensive. An LLM can use
the form of $f$ to propose relevant symbolic structures; \method{} therefore
leaves coefficient fitting and candidate validation to optimization. Because
the structure controls expressiveness, a missing term can make
\eqref{eq:pd}--\eqref{eq:nd} infeasible, while an oversized structure increases
the fitting cost. The proposer samples structures as
}
\begin{equation}\label{eq:proposer}
\begin{aligned}
S_r&\sim\pi_\theta(\mathord{\cdot}\!\mid\!f), &&r=1,\dots,K,\\[-1pt]
S_r&=\{b_{r1},\dots,b_{rm_r}\}\subset\mathcal B,
&&1\le m_r\le m_{\max}.
\end{aligned}
\end{equation}
{%
Here $\theta$ denotes the proposer parameters, $K$ is the number of sampled
structures, and $m_r=\lvert S_r\rvert$. Since the model operates on text,
$f$ and $S$ enter $\pi_\theta$ as strings, in the format shown in Fig.~\ref{fig:interface}. Equivalent structures are
canonicalized and deduplicated before fitting, so repeated samples do not
consume the proposal budget.
}

\begin{figure}[t]
\centering
\input{fig_interface.tex}
\caption{Proposer interface. The prompt is a shortened excerpt of the stored messages; the output is the structure of the candidate $\tfrac16x_1(e^{x_1}-1)+x_2^2$ recovered for that system.}
\label{fig:interface}
\end{figure}
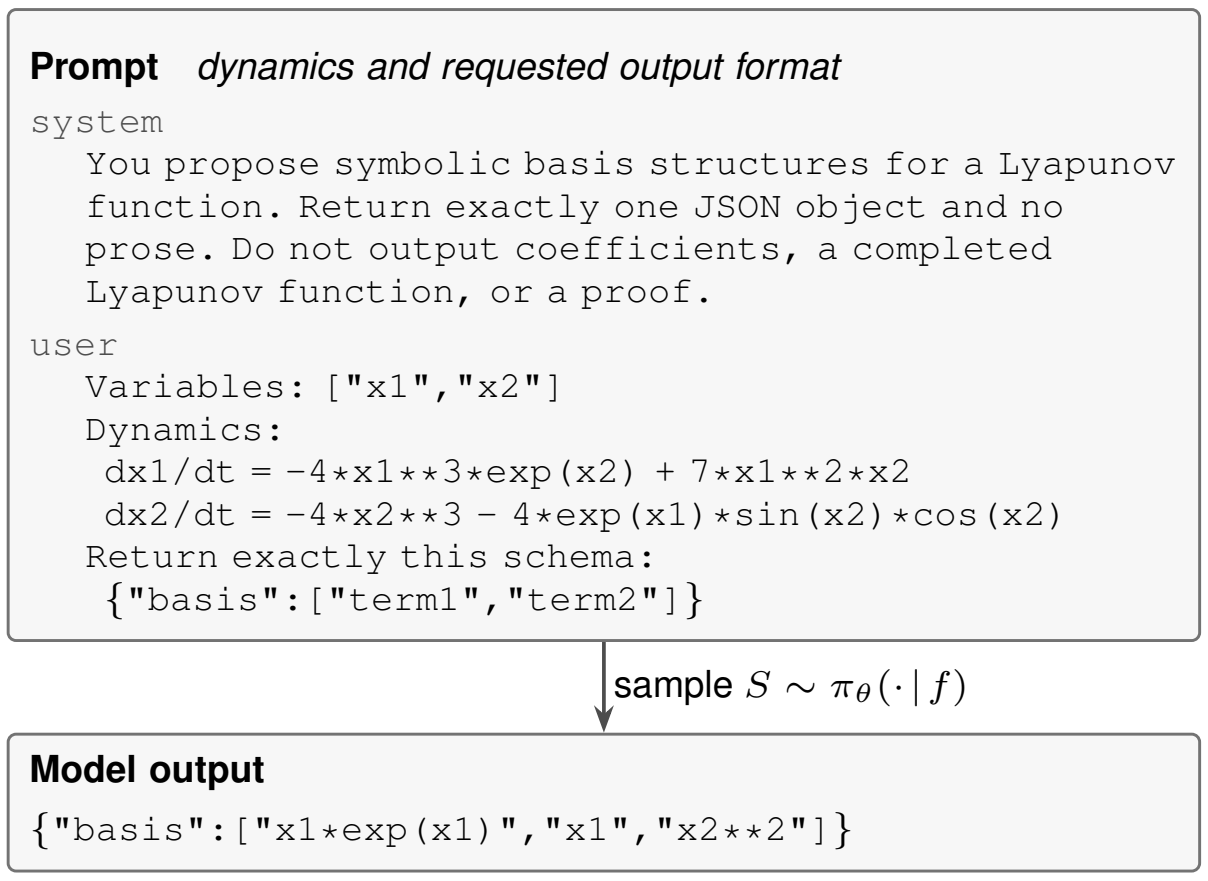

{
When none of the $K$ proposals passes validation, local structure search
provides a deterministic fallback. It starts from the most promising proposals
and builds a finite pool $\mathcal B_f\subset\mathcal B$ from their terms,
monomials of total degree two through four, and simple products involving
elementary transcendental expressions already present in $f$ (such as
$\exp$, $\sin$, or $\cos$), optionally multiplied by a state variable or its
square. The pool is deduplicated and capped. Each step adds, deletes, or
exchanges one term, refits the coefficients under a common fitting set and
budget, and keeps an edit only when it strictly improves the recorded score.
}

\subsection{Training the Proposer}\label{sec:training}
The proposer is trained in two stages. Stage~1 learns the mapping from dynamics
to structures using solved examples, while Stage~2 adds supervision for
structures recovered on systems that Stage~1 misses. This feedback loop is the
core of our training procedure.

\emph{Stage 1: training on known structures.}
A pretrained model may encode general Lyapunov concepts without a
system-specific mapping to valid structures. Stage~1 learns this mapping from
pairs of dynamics and known structures. We discard scalar multipliers and
retain only the terms, denoted by $\operatorname{basis}(V)$, as in
\[
\begin{aligned}
V&=2x_1^2+2x_1x_2+x_2^2+x_3^2,\\
\operatorname{basis}(V)&=\{x_1x_2,x_1^2,x_2^2,x_3^2\}.
\end{aligned}
\]
Each row pairs the serialized dynamics $q=\sigma(f)$ with the serialized
structure $y=\sigma(S)$. Removing coefficients focuses training on structure
selection rather than numerical fitting.

From $\theta_0$, Stage~1 minimizes the autoregressive supervised fine-tuning
(SFT) loss on $\mathcal D$, which combines data from Alfarano et
al.~\cite{alfarano2024global} with additional generated systems.
For backward-generated examples, the generator first constructs $V$
and then generates dynamics satisfying $\nabla V^\top f\le0$. The objective is
\begin{equation}\label{eq:sft}
\mathcal L_1(\theta)=-\mathbb E_{(q,y)\sim\mathcal D}\left[\sum_{j=1}^{|y|}\log\pi_\theta(y_j\mid q, y_{<j})\right],
\end{equation}
where $y_{<j}=(y_1,\dots,y_{j-1})$ denotes the tokens preceding $y_j$. One epoch with low-rank adapters~\cite{hu2022lora} produces $\theta_1$.

\emph{Stage 2: training on search-recovered structures.}
Stage~1 only covers structures in $\mathcal D$. We therefore run it on a fresh,
benchmark-disjoint pool $\mathcal P$ generated separately.
For systems with no accepted proposal, local search recovers structures
$\mathcal R_f$; the corresponding pairs $(q,y)=(\sigma(f),\sigma(S))$ form
$\mathcal F$. Stage~2 continues SFT from $\theta_1$ on the mixed corpus
$\mathcal D_W$: $\mathcal D$ plus $W$ copies of each row in $\mathcal F$ (total $|\mathcal D|+W|\mathcal F|$ rows).
\begin{equation}\label{eq:sft2}
\mathcal L_2(\theta)=-\mathbb E_{(q,y)\sim\mathcal D_W}
\left[\sum_{j=1}^{|y|}\log\pi_\theta(y_j\mid q,y_{<j})\right].
\end{equation}
The expectation is over these $|\mathcal D|+W|\mathcal F|$ rows.
Algorithm~\ref{alg:train} summarizes the loop; the fitting loss
$\ell_S(\cdot;X)$, fitted coefficients $\widehat c_S$, and attained value
$\widehat J_X$ are defined in Section~\ref{sec:coeff}.

\begin{algorithm}[t]
\caption{Search-Augmented Proposer Training.}
\label{alg:train}
\begin{algorithmic}[1]
\REQUIRE pretrained $\theta_0$; corpus $\mathcal D=\{(f,S)\}$; fresh pool $\mathcal P$, disjoint from $\mathcal D$ and from every benchmark; learning rate $\eta$; $T$ steps per stage on weights $w_t$
\ENSURE structure proposer $\pi_{\theta_2}$
\item[\textbf{{Stage 1}}]
\STATE $w_0\!\leftarrow\!\theta_0$, $w_{t+1}\!\leftarrow\! w_t-\eta\nabla\mathcal L_1(w_t)$, $\theta_1\!\leftarrow\! w_T$\hfill\eqref{eq:sft}
\item[\textbf{{Stage 2}}]
\STATE $\mathcal F\leftarrow\varnothing$
\FOR{each system $f\in\mathcal P$}
\STATE $S_1,\dots,S_K\sim\pi_{\theta_1}(\mathord{\cdot}\!\mid\!f)$\hfill\eqref{eq:proposer}
\STATE fit $\widehat c_{S_r}$ using $\ell_{S_r}(\cdot;X)$ and normalize\hfill\eqref{eq:jx}
\STATE validate each $V_{S_r,\widehat c_{S_r}}(x)-V_{S_r,\widehat c_{S_r}}(0)$\hfill\eqref{eq:pd}--\eqref{eq:nd}
\IF{no proposal is accepted}
\STATE $\mathcal R_f\leftarrow$ local search from the proposals ranked best by $\widehat J_X$
\STATE $\mathcal F\leftarrow\mathcal F\cup\{(f,S):S\in\mathcal R_f\}$
\ENDIF
\ENDFOR
\STATE $w_0\!\leftarrow\!\theta_1$, $w_{t+1}\!\leftarrow\! w_t-\eta\nabla\mathcal L_2(w_t)$, $\theta_2\!\leftarrow\! w_T$\hfill\eqref{eq:sft2}
\end{algorithmic}
\end{algorithm}

\subsection{Coefficient Optimization}\label{sec:coeff}
Given $S$, we choose coefficients $c$ so that the candidate satisfies the
Lyapunov conditions. By \eqref{eq:cand}, the constraints are affine in $c$;
the continuous-domain problem is a semi-infinite LP, which we approximate by
sampling $\Omega$~\cite{sankaranarayanan2013handelman,bensassi2016linear,hettich1993semiinfinite}.
For fixed $S$, the sampled problem is convex in $c$; the combinatorial search
is therefore concentrated on selecting $S$, while full-domain validation
remains a separate step.

Let $X=\{x^{(1)},\dots,x^{(N)}\}\subset\Omega_\epsilon$, where
$\Omega_\epsilon=\{x\in\Omega:\norm{x}\ge\epsilon\}$. For fixed $S$, define
$V_c(x)=V_{S,c}(x)-V_{S,c}(0)$, so that $V_c(0)=0$ and
$\Vdot_c(x)=\Vdot_{S,c}(x)$. For coefficient fitting, we impose the Lyapunov
inequalities at sampled points and use nonnegative slack variables $u_i$ and
$v_i$ to measure violations. The inner radius avoids imposing
a quadratic margin arbitrarily close to the origin; validation uses
\eqref{eq:pd}--\eqref{eq:nd} on $\Omega$.
\begin{equation*}
\begin{aligned}
\min_{c,u,v}\quad
&\sum_{i=1}^{N}(u_i+v_i)\\[-1pt]
\text{s.t.}\quad
&V_c(x^{(i)})\ge\gamma\norm{x^{(i)}}^{2}-u_i,\\[-1pt]
&\Vdot_c(x^{(i)})\le-\gamma\norm{x^{(i)}}^{2}+v_i,\\[-1pt]
&u_i,v_i\ge0,\qquad\norm{c}_\infty\le1.
\end{aligned}
\end{equation*}
Here $u_i,v_i$ are pointwise violations, in contrast to the uniform bounds
$u,v$ in \eqref{eq:nestedsearch}. Eliminating these slacks gives a hinge loss;
we use its mean form to normalize the score:
\begin{equation}\label{eq:jx}
\begin{aligned}
\ell_S(c;X):={}&\frac1N\sum_{i=1}^{N}\Big([\gamma\norm{x^{(i)}}^{2}-V_c(x^{(i)})]_+\\[-1pt]
&\qquad\ +[\gamma\norm{x^{(i)}}^{2}+\Vdot_c(x^{(i)})]_+\Big),\\[2pt]
J_X(S):={}&\min_{\norm{c}_\infty\le1}\ell_S(c;X),
\end{aligned}
\end{equation}
where $[z]_+:=\max\{z,0\}$.

We numerically minimize $\ell_S(c;X)$ subject to the box constraint
$\norm{c}_\infty\le1$, then normalize and round the coefficients to obtain
$\widehat c_S$, which need not attain $J_X(S)$.
We rank structures by
$\widehat J_X(S)=\ell_S(\widehat c_S;X)$ and validate the shifted candidate
$V_{S,\widehat c_S}(x)-V_{S,\widehat c_S}(0)$.
Fitting samples do not cover all of $\Omega$; validation checks candidates on $\Omega$. Terms retain canonical scales, so scores depend on basis scale.

\section{Evaluation on Literature and Synthetic Benchmarks}\label{sec:results}

\begin{table*}[t]
\caption{Benchmark comparison of accepted systems across methods.}
\label{tab:b600}
\centering
\footnotesize
\setlength{\tabcolsep}{6.84pt}
\renewcommand{\arraystretch}{1.2}
\begin{tabular}{@{}lccccccc@{\quad}c@{}}
\toprule
& & \multicolumn{3}{c}{\textbf{Neural}} & \multicolumn{2}{c}{\textbf{SOS}}
& \textbf{Generative} & \multirow{2}{*}[-2pt]{\shortstack{\textbf{\method}\\[-1pt]\textbf{(Ours)}}}\\
\cmidrule(r){3-5}\cmidrule(lr){6-7}\cmidrule(l){8-8}
\textbf{Test set} & \textbf{Size} & \textsc{ANLC}~\cite{grande2023anlc} &
\textsc{Fossil}~\cite{edwards2024fossil} & \textsc{LyzNet}~\cite{lyznet2024} &
\texttt{findlyap}~\cite{alfarano2024global} & \textsc{SOSTOOLS}~\cite{prajna2002sostools} &
transformer~\cite{alfarano2024global} & \\
\midrule
Synthetic NonPoly & $200$ & $34$\,($17\%$) & $49$\,($25\%$) & $19$\,($10\%$) & --- & ---
& $60$\,($30\%$) & $\mathbf{108}$\,($\mathbf{54\%}$)\\
Synthetic Poly3 & $200$ & $23$\,($12\%$) & $15$\,($8\%$) & $5$\,($3\%$) & $45$\,($23\%$)
& $85$\,($43\%$) & $43$\,($22\%$) & $\mathbf{111}$\,($\mathbf{56\%}$)\\
Synthetic Poly5 & $200$ & $20$\,($10\%$) & $8$\,($4\%$) & $4$\,($2\%$) & $47$\,($24\%$)
& $89$\,($45\%$) & $38$\,($19\%$) & $\mathbf{107}$\,($\mathbf{54\%}$)\\
Synthetic, all & $600$ & $77$\,($13\%$) & $72$\,($12\%$) & $28$\,($5\%$) & $92$\,($15\%$)
& $174$\,($29\%$) & $141$\,($24\%$) & $\mathbf{326}$\,($\mathbf{54\%}$)\\
Control literature & $109$ & $63$\,($58\%$) & $50$\,($46\%$) & $44$\,($40\%$) & $51$\,($47\%$)
& $68$\,($62\%$) & $41$\,($38\%$) & $\mathbf{92}$\,($\mathbf{84\%}$)\\
\bottomrule
\end{tabular}

\vspace{2pt}
{\footnotesize\raggedright A dash marks a setting the method does not support; percentages are over the full test set.\par}
\end{table*}

\begin{figure*}[t]
\vspace*{6pt}
\centering
\includegraphics[width=\textwidth]{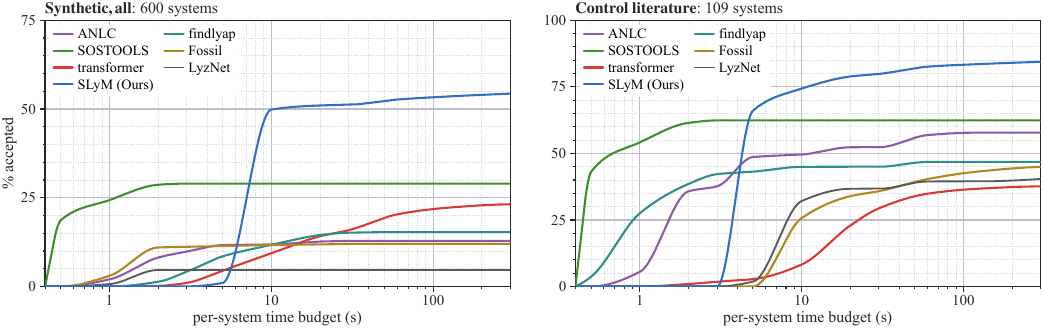}
\caption{Accepted-system rate versus per-system time budget on the synthetic (left) and literature (right) benchmarks. The budget is measured at inference and covers generation, coefficient optimization, validation, and local structure search; one-time training is excluded for \method{} and for the pretrained transformer alike.}
\label{fig:budget}
\end{figure*}

\subsection{Evaluation Setup}\label{sec:setup}

\textbf{Benchmarks.} We evaluate on two benchmarks. The literature benchmark contains $109$ published control examples with declared domains and reference Lyapunov functions; these systems are held out from both $\mathcal D$ and $\mathcal P$. The synthetic benchmark contains $600$ generated systems, equally divided among three families from Alfarano et al.~\cite{alfarano2024global} and labeled below: \emph{NonPoly} is the non-polynomial family of two or three equations, \emph{Poly3} the polynomial family of two or three equations, and \emph{Poly5} the polynomial family of two to five equations. We follow their data-generation and split protocol, and exclude any system whose canonical dynamics appear in a training corpus. The split follows the instance-level protocol of Alfarano et al.

\textbf{Baselines.} We compare with three neural methods, \textsc{ANLC}~\cite{grande2023anlc}, \textsc{Fossil}~\cite{edwards2024fossil}, and \textsc{LyzNet}~\cite{lyznet2024}; two SOS implementations, \texttt{findlyap}~\cite{alfarano2024global} and \textsc{SOSTOOLS}~\cite{prajna2002sostools}; and the symbolic transformer of Alfarano et al.~\cite{alfarano2024global}. We evaluate eight \textsc{ANLC} configurations and run \textsc{SOSTOOLS} at degrees $2$, $4$, and $6$, validating every feasible solution.

\textbf{Common protocol.} All methods receive the same $300$\,s per-system budget on the same machine, spent across their attempted degrees or configurations as each sees fit. Whatever candidates a method returns within that budget are judged by a single validation pipeline for \eqref{eq:pd}--\eqref{eq:nd}, using interval/SMT checks, dense sampling, numerical falsification, and an origin-neighborhood routine. Acceptance uses the full pipeline; counts include formal and numerical screening. A system is accepted once one of its candidates clears the pipeline; anything else, a timeout included, is a failure.

\textbf{\method{} configuration.} The proposer is Qwen3.5-2B~\cite{qwen35}, sampling $K=50$ structures of at most $m_{\max}=10$ terms, with powers up to $p_{\max}=12$ and at most $80$ operations per term; local structure search runs from at most eight seeds over a pool $\mathcal B_f$ capped at $60$ terms. Coefficient optimization uses $\gamma=10^{-2}$ and $\epsilon=0.3$ unless the instance declares another inner radius, fitting on the subset of $2{,}000$ points drawn from the enclosing box that lie in $\Omega_\epsilon$, corner-enriched on box domains; refits inside local search use a smaller subsample. Each stage is one epoch of LoRA fine-tuning (rank $16$, scale $32$, dropout $0.05$, batch size $16$; Stage~1 learning rate $3\times10^{-5}$ and sequence length $768$; Stage~2: $2\times10^{-5}$ and $512$) on the $\lvert\mathcal D\rvert=30{,}000$ known pairs, with Stage~2 including each of the $\lvert\mathcal F\rvert=820$ recovered rows $W=10$ times. Training took $7.4$ GPU-hours on one NVIDIA RTX 4090.

\subsection{Main Results and Solve Time}\label{sec:main}

Under the common acceptance criterion, \method{} accepts $92/109$ systems on the literature benchmark versus $68/109$ for \textsc{SOSTOOLS} (Table~\ref{tab:b600}), with accepted structures that differ from the published forms, as illustrated in Table~\ref{tab:examples}, where coefficients are shown rounded and the rounded form is the one validated. On the synthetic benchmark, generated as in~\cite{alfarano2024global}, it accepts $326/600$ versus $174/600$ for \textsc{SOSTOOLS} ($174/400$ on the polynomial strata it supports, against our $218/400$) and $141/600$ for the symbolic transformer. It leads each synthetic stratum, including $108/200$ non-polynomial systems outside the SOS setting.

\begin{table}[t]
\caption{Representative Lyapunov functions discovered by \method.}
\label{tab:examples}
\centering
\begin{tikzpicture}[
  card/.style={
    draw=black!25,
    fill=black!1.5,
    rounded corners=3pt,
    line width=0.55pt,
    inner sep=4.5pt,
    text width=\columnwidth-11pt,
    align=left
  }
]

\node[card] (c1) {
  \textbf{Region-of-attraction example}~\cite{anderson2015advances}\\[-3pt]
  {\color{black!15}\rule{\linewidth}{0.4pt}}\\[2pt]
  {\small\(\displaystyle \dot x_1=x_2,\qquad \dot x_2=-x_1+\tfrac13x_1^{3}-x_2\)}\\[2pt]
  {\scriptsize\emph{published:}\ \(\displaystyle V(x)=3.421x_1^{2}+1.7217x_1x_2+2.8584x_2^{2}+0.45219x_1^{4}\)}\\[0.5pt]
  {\scriptsize\(\displaystyle \hphantom{V(x)={}}+\,1.318x_1^{3}x_2+1.5945x_1^{2}x_2^{2}+0.20294x_1x_2^{3}+0.86584x_2^{4}\)}\\[1.5pt]
  {\footnotesize\method{}:\ \(\displaystyle V(x)=x_1^{2}+x_1x_2+x_2^{2}\)}\\[1pt]
  {\scriptsize Both hold on \(x^{\top}x\le2.2\)~\cite{anderson2015advances}.}
};

\node[card, below=4pt of c1] (c2) {
  \textbf{Krasovskii example}~\cite{slotine1991applied}\\[-3pt]
  {\color{black!15}\rule{\linewidth}{0.4pt}}\\[2pt]
  {\small\(\displaystyle \dot x_1=-6x_1+2x_2,\qquad \dot x_2=2x_1-6x_2-2x_2^{3}\)}\\[2pt]
  {\footnotesize\emph{published:}\ \(\displaystyle V(x)=(-6x_1+2x_2)^{2}+(2x_1-6x_2-2x_2^{3})^{2}\)}\\[1.5pt]
  {\footnotesize\method{}:\ \(\displaystyle V(x)=x_1^{4}+x_2^{4}\)}
};
\end{tikzpicture}
\end{table}

Fig.~\ref{fig:budget} shows the coverage--time trade-off: SOS reaches its synthetic total within $3$\,s, while \method{} exceeds every evaluated baseline's final coverage within $4.7$\,s (literature) and $7.4$\,s (synthetic), both including generation. It finds $299/326$ synthetic successes within $10$\,s. Proposals are consumed sparingly: on $170$ of the $326$ accepted systems the first proposal already succeeds. Local structure search contributes $21$ of the total, evaluating a median of $217$ edited structures and $3$ validation calls per system, at a median post-generation cost of $2.34$\,s.

\subsection{Ablation: Decomposition and Training}\label{sec:ablation}
In Table~\ref{tab:ablate}, moving from \emph{complete $V$} to \emph{structures} roughly doubles coverage across all proposers under fixed budgets: $64\to126$ for the base model, $153\to297$ after Stage~1, and $141\to263$ for the symbolic transformer of Alfarano et al.~\cite{alfarano2024global}. Transformer terms are refit with the common coefficient optimizer.

Training the proposer adds further gains, especially where the base model is weakest. Moving from the base model to Stage~1 lifts synthetic coverage from $126$ to $297$ (Poly5: $34\to99$, NonPoly: $47\to91$) and literature coverage from $78$ to $85$. The final pipeline uses Stage~2 search-augmented supervision and reaches the $326$ synthetic and $92$ literature successes reported in Table~\ref{tab:b600}.

Permutation and permutation-plus-rescaling reruns yield $91$ and $90$ successes, respectively, versus $91/109$ for the matched untransformed control, probing sensitivity to equivalent coordinate representations. Sweeping budget $K$ shows that the pipeline accepts $276$ synthetic systems even with one initial proposal ($K=1$), followed by fitting and local search, and plateaus beyond $K=25$ (Fig.~\ref{fig:ksweep}), while Stage-1 proposals vary by only $\pm3$ across four generation seeds.

\begin{table}[t]
\caption{Ablations at $K=50$.}
\label{tab:ablate}
\centering
\scriptsize
\setlength{\tabcolsep}{2.6pt}
\begin{tabular*}{\columnwidth}{@{\extracolsep{\fill}}lccccc@{}}
\toprule
& \multicolumn{4}{c}{\textsc{Synthetic}} & \\
\cmidrule(lr){2-5}
\textbf{Configuration} & \textbf{NonPoly} & \textbf{Poly3} & \textbf{Poly5}
& \textbf{all} & \textbf{Literature}\\
& $200$ & $200$ & $200$ & $600$ & $109$\\
\midrule
Base model, complete $V$ & $20$ & $23$ & $21$ & $64$ & $47$\\
Base model, structures   & $47$ & $45$ & $34$ & $126$ & $78$\\
Stage 1, complete $V$    & $59$ & $49$ & $45$ & $153$ & $73$\\
Stage 1, structures      & $91$ & $107$ & $99$ & $297$ & $85$\\
Transformer~\cite{alfarano2024global}, complete $V$ & $60$ & $43$ & $38$ & $141$ & $41$\\
Transformer~\cite{alfarano2024global}, structures & $87$ & $93$ & $83$ & $263$ & $59$\\
\textbf{\method} & $\mathbf{108}$ & $\mathbf{111}$ & $\mathbf{107}$
& $\mathbf{326}$ & $\mathbf{92}$\\
\bottomrule
\end{tabular*}

\vspace{2pt}
{\scriptsize\raggedright \emph{Complete $V$}: the proposer emits a full function with coefficients. Stage~1 complete-$V$ uses full-function targets; structure rows use term-only targets. Base model: Qwen3.5-2B without fine-tuning.\par}
\end{table}

\begin{figure}[t]
\centering
\includegraphics[width=\columnwidth]{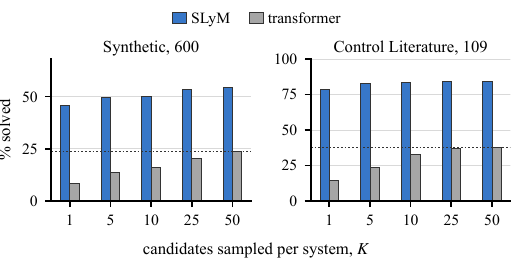}
\caption{Acceptance versus initial proposal budget $K$. \method{} runs coefficient optimization and local structure search after its $K$ structure proposals; the external transformer~\cite{alfarano2024global} generates $K$ complete functions. The dashed line marks that transformer's result at $K=50$.}
\label{fig:ksweep}
\end{figure}

\section{Conclusion}\label{sec:conclusion}
We presented \method{}, which discovers symbolic Lyapunov functions by decomposing the construction into structure proposal and coefficient optimization: the LLM proposes a coefficient-free structure beyond rigid fixed bases, a convex program supplies the weights, and search-recovered structures folded back into supervision extend the proposer beyond its training corpus. Across all benchmarks \method{} outperforms the evaluated neural, sum-of-squares, and generative baselines, including on the non-polynomial systems outside the SOS setting. Future work will extend the structure library and search procedure to more complex systems for which the current pipeline finds no candidate.

\newpage
\scriptsize
\bibliographystyle{IEEEtran}
\bibliography{references}

%% file: fig_pipeline.tex
\begin{tikzpicture}[
  x=1mm, y=1mm, font=\sffamily\small,
  >={Stealth[length=1.65mm,width=1.2mm]},
  card/.style={rounded corners=3pt, line width=.55pt,
    minimum width=54mm, minimum height=11mm, inner sep=0pt},
  structure/.style={card, draw=axblue!45, fill=axblue!5},
  coefficient/.style={card, draw=axgreen!50, fill=axgreen!6},
  validation/.style={card, draw=black!28, fill=black!3, minimum height=9mm},
  title/.style={anchor=west, inner sep=0pt, font=\sffamily\bfseries\small,
    text=black!85},
  subtitle/.style={anchor=west, inner sep=0pt, font=\sffamily\scriptsize,
    text=black!70},
  flow/.style={->, line width=.8pt, draw=black!62},
  sflow/.style={flow, draw=axblue!85!black},
  fail/.style={flow, draw=axred!85!black},
  train/.style={->, rounded corners=2pt, line width=.8pt,
    draw=axyellow!65!black, dash pattern=on 2.4pt off 1.8pt},
  edgelab/.style={font=\sffamily\scriptsize, inner sep=1.4pt,
    fill=white, text=black!75},
  icon/.style={line width=.65pt, line cap=round, line join=round}
]

\node[draw=black!25, fill=black!2, rounded corners=7pt,
  minimum width=27mm, minimum height=7mm, inner sep=0pt,
  text=black!80] (sys) at (0,0) {};
\node[inner sep=0pt,text=black!80] at (0,0) {$\dot{x}=f(x)$};
\node[structure] (prop) at (0,-14) {};
\node[coefficient] (fit) at (0,-30) {};
\node[validation] (verify) at (0,-45) {};
\node[structure] (search) at (0,-60) {};

\node[title] at (-15.6,-12.1) {LLM proposer};
\node[subtitle] at (-15.6,-16.1)
  {$K$ samples $S\sim\pi_\theta(\mathord{\cdot}\!\mid\!f)$};
\node[title] at (-15.6,-28.1) {coefficient optimization};
\node[subtitle] at (-15.6,-32.1) {optimize coefficients $c$};
\node[title] at (-15.6,-45) {candidate validation};
\node[title] at (-15.6,-58.1) {local structure search};
\node[subtitle] at (-15.6,-62.1) {edit one symbolic term};

\begin{scope}[shift={(-21.7,-14)}, icon, draw=axblue!90!black]
  \fill[axblue!12] (0,0) circle (3.75);
  \draw (-1.8,1.55)--(.15,.8)--(-1.8,-1.5)--(.9,-1.25)--(.15,.8);
  \draw (-1.8,1.55)--(-1.8,-1.5);
  \foreach \p in {(-1.8,1.55),(.15,.8),(-1.8,-1.5),(.9,-1.25)}
    \filldraw[fill=white] \p circle (.43);
  \path[fill=axblue,draw=none] (1.75,2.65)--(2.08,1.77)--(2.95,1.45)
    --(2.08,1.12)--(1.75,.25)--(1.42,1.12)--(.55,1.45)--(1.42,1.77)--cycle;
\end{scope}

\begin{scope}[shift={(-21.7,-30)}, icon, draw=axgreen!80!black]
  \fill[axgreen!14] (0,0) circle (3.75);
  \draw (-2,1.6)--(2,1.6) (-2,0)--(2,0) (-2,-1.6)--(2,-1.6);
  \filldraw[fill=white] (-.85,1.6) circle (.5);
  \filldraw[fill=white] (.95,0) circle (.5);
  \filldraw[fill=white] (-.3,-1.6) circle (.5);
\end{scope}

\begin{scope}[shift={(-21.7,-45)}, icon, draw=black!62]
  \fill[black!7] (0,0) circle (3.5);
  \draw[rounded corners=.45pt] (-1.6,-2.1) rectangle (1.6,2.1);
  \draw (-.85,1.05)--(.75,1.05);
  \draw[line width=.85pt] (-.85,-.65)--(-.2,-1.3)--(.95,.05);
\end{scope}

\begin{scope}[shift={(-21.7,-60)}, icon, draw=axblue!90!black]
  \fill[axblue!12] (0,0) circle (3.75);
  \draw[line width=.9pt] (-.65,.65) circle (1.7);
  \draw[line width=1.35pt] (.6,-.6)--(2.2,-2.2);
  \draw[line width=.5pt] (-1.4,.25)--(-.65,1.1)--(.1,.25);
  \foreach \p in {(-1.4,.25),(-.65,1.1),(.1,.25)}
    \fill \p circle (.23);
\end{scope}

\draw[flow] (sys.south)--(prop.north);
\draw[sflow] (prop.south)--node[edgelab,right=.4mm] {structure $S$} (fit.north);
\draw[flow] (fit.south)--node[edgelab,right=.4mm] {$V_{S,c}$} (verify.north);
\draw[fail] (verify.south)--node[edgelab,right=.4mm,text=axred!85!black] {fail}
  (search.north);
\draw[sflow,rounded corners=2pt] (search.west)--(-31.5,-60)|-(fit.west);

\node[draw=axblue!65, fill=axblue!10, line width=.75pt, rounded corners=3pt,
  minimum width=17mm, minimum height=11mm, inner sep=0pt,
  align=center, text=axblue!85!black, font=\sffamily\small]
  (acc) at (44.3,-45) {};
\node[inner sep=0pt,text=axblue!85!black,font=\sffamily\bfseries\small]
  at (44.3,-42.8) {\textbf{accepted}};
\node[inner sep=0pt,text=axblue!85!black] at (44.3,-47.1) {$V_{S,c}$};
\draw[flow] (verify.east)--node[edgelab,above,inner sep=.8pt] {pass} (acc.west);

\draw[train] (acc.north)|-(prop.east);
\node[text width=21mm, align=center, fill=white, rounded corners=2pt, inner sep=2pt,
  font=\sffamily\scriptsize, text=axyellow!50!black] at (43.4,-28.9)
  {training feedback};
\end{tikzpicture}

%% file: fig_interface.tex
\begin{tikzpicture}[
  >={Stealth[length=1.8mm,width=1.3mm]},
  panel/.style={draw=black!55, fill=black!3, rounded corners=1.5pt,
                line width=0.55pt, inner sep=4pt, text width=76mm,
                align=left, font=\scriptsize},
  promptp/.style={panel},
  outp/.style={panel},
  arrow/.style={->, draw=black!70, line width=0.7pt},
  arrowlabel/.style={font=\sffamily\scriptsize, inner sep=1.5pt, right}
]

\node[promptp] (prompt) {\\[1pt]
  {\sffamily\bfseries Prompt}\quad{\sffamily\itshape dynamics and requested output
  format}\\[2pt]
  {\ttfamily\raggedright\setlength{\parindent}{0pt}%
   {\color{black!70}system}\par
   \hangindent=1.5em\hangafter=0\relax
   You propose symbolic basis structures for a Lyapunov function. Return exactly one
   JSON object and no prose. Do not output coefficients, a completed Lyapunov
   function, or a proof.\par
   \addvspace{2pt}
   {\color{black!70}user}\par
   \hangindent=1.5em\hangafter=0\relax
   Variables: ["x1","x2"]\par
   \hangindent=1.5em\hangafter=0\relax
   Dynamics:\par
   \hangindent=3.5em\hangafter=1\relax
   \hspace*{2em}dx1/dt = -4*x1**3*exp(x2) + 7*x1**2*x2\par
   \hangindent=3.5em\hangafter=1\relax
   \hspace*{2em}dx2/dt = -4*x2**3 - 4*exp(x1)*sin(x2)*cos(x2)\par
   \hangindent=1.5em\hangafter=0\relax
   Return exactly this schema:\par
   \hangindent=3.5em\hangafter=1\relax
   \hspace*{2em}\{"basis":["term1","term2"]\}%
  }%
};

\node[outp, below=6mm of prompt.south, anchor=north] (out) {%
  {\sffamily\bfseries Model output}\\[3pt]
  {\ttfamily\{"basis":["x1*exp(x1)","x1","x2**2"]\}}%
};

\draw[arrow] (prompt.south) -- (out.north)
  node[arrowlabel, midway] {sample $S\sim\pi_\theta(\mathord{\cdot}\!\mid\!f)$};
\end{tikzpicture}